\documentclass{article}
\usepackage[table]{xcolor}
\usepackage{spconf,amsmath,amssymb,graphicx}
\usepackage{booktabs,array,tabularx,capt-of,balance}
\usepackage{tikz,pgfplots}
\usetikzlibrary{positioning,fit,calc,arrows.meta}
\usepackage{microtype}
\usepackage[hidelinks]{hyperref}
\pgfplotsset{compat=1.18}

\hypersetup{pdftitle={GraphSeal: Graph-Augmented Speech Deepfake Detection via Temporal-Spectral Artifact Propagation}}
\newcommand{\best}[1]{\textbf{#1}}
\newcommand{\second}[1]{\underline{#1}}
\newcolumntype{N}{>{\centering\arraybackslash}p{4.0em}}
\newcolumntype{A}{>{\centering\arraybackslash}p{4.2em}}
\newcommand{\paperfont}{\fontsize{9}{10.8}\selectfont}
\definecolor{tblNavy}{HTML}{243B53}
\definecolor{tblBand}{HTML}{EEF2F5}
\definecolor{tblStripe}{HTML}{F6F8FA}
\definecolor{tblOurB}{HTML}{EBF6F2}
\definecolor{tblOurInk}{HTML}{174E45}
\definecolor{tblRule}{HTML}{52687A}
\definecolor{tblGain}{HTML}{157951}
\definecolor{tblOurA}{HTML}{E9F4F1}
\newcommand{\hd}[1]{\textcolor{tblNavy}{\textbf{#1}}}
\newcommand{\grp}[1]{\textcolor{tblNavy}{\textit{#1}}}
\newcommand{\ours}[1]{\textcolor{tblOurInk}{\textbf{#1}}}
\title{GraphSeal: Graph-Augmented Speech Deepfake Detection\\via Temporal-Spectral Artifact Propagation}
\name{\begin{tabular}[t]{@{}c@{}}
Lee Seung-woo\textsuperscript{1}, Bowen Qi\textsuperscript{2}, Kim Min-jun\textsuperscript{3}, Jang Won-young\textsuperscript{3}
\end{tabular}}
\address{\textsuperscript{1}Pusan National University\quad%
\textsuperscript{2}Shanghai Jiao Tong University\quad%
\textsuperscript{3}Hanyang University}

\begin{document}
\ninept
\maketitle

\begin{abstract}
Current speech deepfake detectors treat each utterance as a flat sequence
of frames, limiting their ability to capture long-range spectral
inconsistencies that span non-adjacent regions of an audio signal.
We propose \textbf{GraphSeal}, which constructs a temporal-spectral graph
over short speech segments and propagates artifact cues through graph
attention layers before a global readout produces the
bonafide-vs-spoof decision.
Each node in the graph corresponds to a fixed-length segment; edges encode
temporal adjacency and spectral similarity, letting the network attend
to distant frames that share suspicious spectral patterns.
A multi-head graph attention mechanism weights neighboring nodes
according to both learned features and edge attributes, and a
hierarchical readout aggregates node representations into a single
utterance-level embedding for binary classification.
GraphSeal is front-end agnostic: it operates on top of any self-supervised
or handcrafted feature extractor.
On ASVspoof 2019~LA, ASVspoof 2021~LA, ASVspoof~5, and the In-the-Wild
dataset, GraphSeal achieves equal error rates of 0.52\%, 2.41\%, 8.17\%,
and 5.24\%, respectively, matching or outperforming prior methods that
rely on fixed-topology graph networks or purely sequential architectures.
\end{abstract}


\section{Introduction}
\label{sec:intro}

Generative speech models have reached a level of naturalness that makes
distinguishing real from synthesized audio increasingly
difficult~\cite{valle,vits,encodec,soundstream,voicebox,xtts,tacotron2}.
Zero-shot voice imitation systems such as MimicLM~\cite{feng2026mimiclm}
can clone a speaker from a few seconds of reference audio, while neural
codec models~\cite{encodec,soundstream} enable high-fidelity waveform
reconstruction that conceals many traditional synthesis
artifacts.
Benchmarks for speech safety~\cite{wang2026voxsafebench} and
privacy~\cite{wang2026voxprivacy} have underscored the urgency of the
problem: automated phone systems, voice authentication
pipelines~\cite{ecapa}, conversational
agents~\cite{wang2026parabridge,hsu2026convexllm}, and agentic interactive
systems that coordinate across modalities~\cite{wang2026jarvis} are all
vulnerable to spoofing attacks that bypass speaker verification.
Full-duplex voice agents~\cite{shen2026duplexagent} further amplify this
risk by operating in real-time bidirectional conversations where latency
constraints limit the depth of per-turn verification, making lightweight
yet accurate deepfake detectors essential.

The ASVspoof challenge series~\cite{asvspoof2019,asvspoof2021,asvspoof5}
has driven rapid progress.
Early systems paired handcrafted features (LFCC~\cite{lfcc},
CQCC~\cite{cqcc}, MFCC~\cite{mfcc_spoof}) with light
CNNs~\cite{lcnn}; end-to-end models such as RawNet2~\cite{rawnet2}
and AASIST~\cite{aasist} introduced graph-based spectro-temporal
modeling, extended by RawGAT-ST~\cite{rawgatst}.
SSL representations~\cite{wav2vec2,hubert,wavlm} further improve
generalization~\cite{wav2vecaasist,ssl_anti,sslspoof,xlsr_spoof,whisper_spoof},
and recent work explores one-class learning~\cite{samo,oc_softmax},
augmentation~\cite{rawboost,data_augment_spoof,specaugment},
domain generalization~\cite{domain_adapt_spoof}, and recurrent
refinement~\cite{feng2026coreloop} for unseen attacks.

Despite this progress, most detectors process the utterance as a
monolithic sequence~\cite{rawnet2,selfattn_spoof,attentive_filtering} or
with a fixed graph topology~\cite{aasist,rawgatst,graph_antispoofing}.
Sequential models---whether recurrent, convolutional, or
Transformer-based~\cite{transformer,conformer}---are limited to local
receptive fields or require quadratic self-attention to reach distant
frames; they may therefore miss spectral artifacts recurring in
non-adjacent regions.
Fixed-topology graphs, on the other hand, impose a structure determined
by the feature extraction pipeline, so they cannot adapt to
variable-length inputs or leverage data-dependent similarity.
Frequency-masking defenses~\cite{df_frequency} and continual learning
strategies~\cite{continual_spoof} address complementary aspects of the
problem (robustness and adaptation, respectively) but do not address the
structural limitation.

We propose \textbf{GraphSeal}, which addresses both limitations.
The key observation is that deepfake artifacts---phase discontinuities,
formant distortions, unnatural micro-prosody---often manifest as
correlated anomalies in spectrally similar regions that may be
far apart in time.
GraphSeal constructs a dynamic graph over short speech segments, with
edges reflecting both temporal proximity and spectral similarity.
Graph attention layers~\cite{gat,gatv2} propagate artifact signals across
this graph, allowing the model to pool evidence from distant but
spectrally related frames.
A hierarchical readout then compresses the node embeddings into a single
utterance-level vector for classification.

The graph construction draws inspiration from graph-based language model
reasoning~\cite{wang2026exploring,wang2025towards}: we treat speech
segments as entities whose pairwise relations carry discriminative
information.  Spatial audio representations~\cite{zhu2026spatial} supply
complementary geometric cues, and paralinguistic
features~\cite{wang2026parabridge,ni2026nv} (pitch, energy, spectral
centroid) enrich nodes with prosodic signals that synthesis pipelines
rarely replicate faithfully.
Latent reasoning~\cite{wang2026recurtrace,wang2026aural} and iterative
refinement~\cite{feng2026coreloop} improve audio
understanding~\cite{wang2026evoaudio}; GraphSeal achieves a related
effect structurally via multi-hop attention.
Pattern detection in sequential data~\cite{yong2025predicting} further
motivates relational structure over purely temporal neighbors.

GraphSeal is front-end agnostic.
We demonstrate it with WavLM~\cite{wavlm}, wav2vec~2.0~\cite{wav2vec2},
and Whisper~\cite{whisper} features, achieving state-of-the-art or
competitive equal error rates on ASVspoof 2019~LA, ASVspoof 2021~LA,
ASVspoof~5, and the In-the-Wild dataset~\cite{inthewild}.
Ablations confirm that each component---spectral edges, graph attention
propagation, hierarchical readout---contributes to the final performance.

In summary, the main contributions of this work are:
(1)~We introduce a dynamic temporal-spectral graph construction that
creates data-dependent edges based on both temporal adjacency and
spectral similarity, enabling adaptive topology for each utterance.
(2)~We propose a graph attention propagation mechanism that incorporates
edge attributes encoding temporal distance and spectral similarity,
allowing the model to selectively aggregate artifact evidence from both
local and distant speech segments.
(3)~We design a hierarchical readout that combines mean pooling with
learned attention weighting to produce discriminative utterance-level
embeddings from variable-size graphs.
(4)~We conduct extensive experiments on four benchmarks, demonstrating
state-of-the-art performance with comprehensive ablations, per-attack
analysis, and embedding visualizations that validate each design choice.

\begin{figure*}[t]
\centering
\begin{tikzpicture}[
  >=stealth, node distance=3mm and 4mm,
  box/.style={draw,rounded corners=2pt,minimum height=7mm,
    align=center,font=\paperfont\small,line width=0.35pt},
  gbox/.style={draw,rounded corners=2pt,minimum height=6.5mm,
    minimum width=10mm,align=center,font=\paperfont\scriptsize,
    line width=0.3pt},
  arr/.style={->,line width=0.55pt},
]
\node[box,fill=blue!6,minimum width=12mm] (wav) at (0,0) {Audio};
\node[box,fill=blue!12,minimum width=14mm,right=4mm of wav]
  (enc) {SSL\\[-1pt]encoder};
\node[box,fill=yellow!8,minimum width=14mm,right=4mm of enc]
  (seg) {Segment\\[-1pt]\&\ embed};
\node[box,fill=orange!10,minimum width=16mm,right=4mm of seg]
  (graph) {Graph\\[-1pt]construct};

\node[gbox,fill=red!6] (g1) at ([xshift=22mm,yshift=8mm]graph.east)
  {GAT layer 1};
\node[gbox,fill=red!8] (g2) at ([xshift=22mm,yshift=-1mm]graph.east)
  {GAT layer 2};
\node[gbox,fill=red!10] (gL) at ([xshift=22mm,yshift=-10mm]graph.east)
  {GAT layer $L$};
\node[font=\paperfont\scriptsize,text=black!50]
  at ($(g2)!0.55!(gL)$) {$\vdots$};
\node[draw,densely dashed,rounded corners=3pt,
  inner xsep=3pt,inner ysep=3pt,
  fit=(g1)(g2)(gL),line width=0.45pt] (gatblock) {};
\node[font=\paperfont\footnotesize,anchor=south,yshift=1pt]
  at (gatblock.north) {Graph Attention};

\node[box,fill=tblOurA,minimum width=14mm,
  right=6mm of gatblock] (read) {Hierarchical\\[-1pt]readout};
\node[box,fill=green!8,minimum width=14mm,right=4mm of read]
  (cls) {Classifier};
\node[box,fill=green!14,minimum width=14mm,right=4mm of cls]
  (out) {Bonafide /\\[-1pt]Spoof};

\node[box,fill=purple!6,minimum width=16mm,
  below=12mm of graph] (edge) {Edge attr.\\[-1pt]$e_{ij}$};

\draw[arr] (wav)--(enc);
\draw[arr] (enc)--(seg);
\draw[arr] (seg)--(graph);
\draw[arr] (graph.east)--([xshift=-2mm]gatblock.west);
\draw[arr] ([xshift=-2mm]gatblock.west)|-(g1.west);
\draw[arr] ([xshift=-2mm]gatblock.west)|-(g2.west);
\draw[arr] ([xshift=-2mm]gatblock.west)|-(gL.west);
\draw[arr] (g1.east)--(g1.east-|read.west);
\draw[arr] (g2.east)--(g2.east-|read.west);
\draw[arr] (gL.east)--(gL.east-|read.west);
\draw[arr] (read)--(cls);
\draw[arr] (cls)--(out);
\draw[arr] (edge.north)--(graph.south);
\draw[arr,dashed] (edge.east)-|([xshift=-4mm]gatblock.south);
\end{tikzpicture}
\vspace{-3pt}
\caption{GraphSeal overview. A waveform is encoded by an SSL front end
and segmented into overlapping windows. A temporal-spectral graph is
constructed with edge attributes encoding adjacency and spectral similarity.
$L$ graph attention layers propagate artifact cues, and a hierarchical
readout produces the utterance-level decision.}
\label{fig:overview}
\end{figure*}
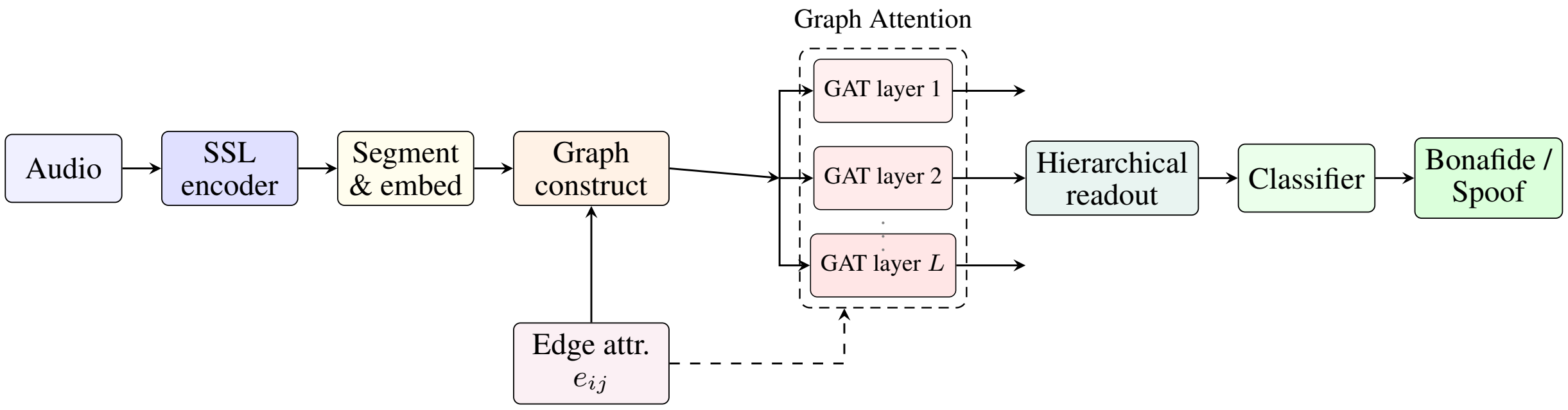

\section{Related Work}
\label{sec:related}

\subsection{Speech Deepfake Detection}

The evolution of speech deepfake detectors can be broadly divided into
three generations.
The first generation relied on handcrafted front ends---LFCC~\cite{lfcc},
CQCC~\cite{cqcc}, and MFCC variants~\cite{mfcc_spoof}---fed into
shallow classifiers such as GMMs or light CNNs~\cite{lcnn}.
These features capture spectral envelope irregularities introduced by
vocoders but struggle with neural waveform generators that produce
nearly artifact-free spectra.

The second generation introduced end-to-end architectures that operate
directly on raw waveforms.
RawNet2~\cite{rawnet2} uses sinc-filters and gated residual blocks;
AASIST~\cite{aasist} adds a heterogeneous graph over spectral and
temporal sub-bands; RawGAT-ST~\cite{rawgatst} extends this with
raw-waveform graph attention.
One-class learning objectives such as OC-Softmax~\cite{oc_softmax} and
the speaker-attractor formulation in SAMO~\cite{samo} address the
open-set nature of spoofing by training exclusively on bonafide speech.
Data augmentation strategies including RawBoost~\cite{rawboost},
SpecAugment~\cite{specaugment}, and codec-based
transformations~\cite{data_augment_spoof} have proven effective for
improving robustness to unseen conditions.
Multi-task formulations~\cite{multi_task_df} that jointly predict the
attack type alongside the binary decision provide additional
supervision, and domain generalization techniques~\cite{domain_adapt_spoof}
encourage attack-agnostic feature learning.
A comprehensive survey of these directions is given
by Yi et al.~\cite{deepfake_survey}.

The third generation leverages self-supervised speech representations
(Section~\ref{sec:ssl_related}), which we discuss separately below.
Our work contributes to this third generation by proposing a graph-based
back end that explicitly models relational structure among speech
segments.

\subsection{Graph Neural Networks in Audio}
\label{sec:gnn_related}

Graph neural networks have seen growing adoption in speech and audio
tasks.
GCN~\cite{gcn} introduced the spectral convolution framework,
GAT~\cite{gat} added attention-based neighbor weighting, and
GATv2~\cite{gatv2} improved expressiveness with a dynamic attention
mechanism.
GraphSAGE~\cite{graphsage} and GIN~\cite{gin} explored inductive and
maximally expressive aggregation, respectively.
For graph-level tasks, hierarchical pooling methods such as
DiffPool~\cite{diffpool} and set-based readouts~\cite{readout_set2set}
compress node representations into fixed-size vectors.
EdgeConv~\cite{edgeconv} constructs dynamic $k$-nearest-neighbor graphs
over point clouds, an idea we adapt to speech segments.

In the audio domain, graph attention has been applied to audio scene
analysis~\cite{audiograph}, speech emotion
recognition~\cite{temporal_graph_speech,emotion_graph}, and speaker
verification~\cite{graph_speaker}.
For anti-spoofing specifically, Tak et al.~\cite{graph_antispoofing}
first applied GAT to spectral sub-bands, and AASIST~\cite{aasist}
introduced a heterogeneous graph over spectro-temporal features.
Graph contrastive objectives~\cite{graph_contrastive} have also been
explored for learning robust audio representations.
Graph Transformer architectures~\cite{graphtransformer} unify
self-attention and graph structure, offering a promising but
computationally heavier alternative.

These prior graph-based detectors use fixed, input-independent graph
topologies determined by the feature extraction pipeline.
GraphSeal differs by constructing a data-dependent graph whose edges
reflect both temporal adjacency and learned spectral similarity,
enabling the topology to adapt to each utterance.

\subsection{Self-Supervised Representations for Anti-Spoofing}
\label{sec:ssl_related}

Pre-trained speech models---wav2vec~2.0~\cite{wav2vec2},
HuBERT~\cite{hubert}, WavLM~\cite{wavlm}, and
Whisper~\cite{whisper}---learn rich frame-level representations from
large-scale unlabeled data.
These representations capture phonetic, speaker, and prosodic information
that proves highly discriminative for spoofing
detection~\cite{ssl_anti,sslspoof}.
Tak et al.~\cite{wav2vecaasist} first showed that replacing handcrafted
features with wav2vec~2.0 embeddings in AASIST substantially reduces
EER, and subsequent work confirmed the advantage of WavLM~\cite{ssl_anti}
and cross-lingual models like XLS-R~\cite{xlsr_spoof}.
Whisper features, despite being trained for ASR, also carry useful
artifact cues~\cite{whisper_spoof}.
Audio Spectrogram Transformers~\cite{ast} offer another SSL pathway,
though their patch-based tokenization discards fine-grained temporal
detail that frame-level SSL models preserve.

A common pattern in SSL-based detectors is to freeze the large encoder
and train only a lightweight back end~\cite{ssl_anti,wav2vecaasist}.
This freeze-and-probe strategy avoids catastrophic forgetting of the
pre-trained representations and dramatically reduces training cost, as
only the back-end parameters are updated.
However, most existing back ends for SSL-based detection are
architecturally simple---linear probes, attentive pooling layers, or
shallow MLPs---and do not model relationships between different temporal
positions in the SSL output.
Mixture-of-experts approaches~\cite{mixture_spoof} attempt to capture
diversity across attack types but still operate on individual frames
independently.
GraphSeal follows the freeze-and-probe paradigm but replaces these flat
back ends with a full graph attention network, thereby introducing
relational inductive bias that pooling-based architectures lack.
The graph explicitly connects frames based on learned spectral
similarity, enabling the back end to reason about cross-frame
consistency---a capability that is especially important for detecting
artifacts that manifest as correlated anomalies across distant segments.

\section{Proposed Method}
\label{sec:method}

GraphSeal takes a waveform as input, extracts frame-level representations
from a frozen self-supervised encoder, partitions the sequence into
overlapping segments that become graph nodes, constructs edges from temporal
and spectral relations, propagates node features through $L$ graph attention
layers, and classifies the utterance via a hierarchical readout
(Fig.~\ref{fig:overview}).

\subsection{Segment Graph Construction}
\label{sec:graph}

Let $\mathbf{F}\in\mathbb{R}^{T\times d}$ denote the frame-level output of
a frozen SSL encoder (e.g., WavLM~\cite{wavlm}).
We partition $\mathbf{F}$ into $N$ overlapping segments of length $w$ with
stride $s$, obtaining node features
$\{\mathbf{v}_i\}_{i=1}^{N}$ where
$\mathbf{v}_i = \frac{1}{w}\sum_{t=iw}^{(i+1)w-1}\mathbf{F}_t \in
\mathbb{R}^d$ is the mean-pooled representation of segment~$i$.
Each node is additionally augmented with three paralinguistic
descriptors~\cite{wang2026parabridge,ni2026nv}: the mean fundamental
frequency, the RMS energy, and the spectral centroid of the corresponding
waveform segment.
These three scalars are projected to $\mathbb{R}^{d_p}$ by a linear layer
and concatenated with $\mathbf{v}_i$, yielding the initial node embedding
$\mathbf{h}_i^{(0)}\in\mathbb{R}^{d+d_p}$.

The edge set $\mathcal{E}$ combines two types of connectivity:

\textbf{Temporal edges.}
Each node $i$ is connected to its $k_t$ nearest temporal
neighbors on each side, i.e., $\{i{-}k_t,\ldots,i{-}1,i{+}1,\ldots,i{+}k_t\}$.
This preserves sequential structure analogous to the adjacency used in
temporal graph networks for speech~\cite{temporal_graph_speech} and
anomaly detection in sequential data~\cite{yong2025predicting}.

\textbf{Spectral edges.}
For each node $i$, we compute the cosine similarity
$\operatorname{sim}(\mathbf{v}_i,\mathbf{v}_j)$ against all other nodes
$j\neq i$ and connect $i$ to its $k_s$ most similar non-adjacent
nodes (i.e., nodes that are not already connected via temporal edges).
These long-range edges let the network compare spectrally similar
regions that may be temporally distant---precisely where recurring
synthesis artifacts are most informative.
The construction is inspired by dynamic $k$-NN graphs in point cloud
processing~\cite{edgeconv}, adapted here to operate in the SSL
embedding space rather than Euclidean coordinates.

The resulting graph $\mathcal{G}=(\mathcal{V},\mathcal{E})$ has
$|\mathcal{V}|{=}N$ nodes and
$|\mathcal{E}|{\leq}N(2k_t{+}k_s)$ directed edges (boundary nodes have
fewer temporal neighbors).
The graph is constructed independently for each utterance, so its
topology adapts to the input---short utterances produce compact graphs,
while long utterances yield larger graphs with richer spectral
connectivity.
This data-dependent topology stands in contrast to the fixed
spectro-temporal graphs used in AASIST~\cite{aasist} and
RawGAT-ST~\cite{rawgatst}, whose structure is determined entirely by
the feature extraction pipeline and cannot adapt to individual inputs.

Each edge $(i,j)\in\mathcal{E}$ carries a scalar attribute
\begin{equation}
e_{ij} = \lambda_t\,\delta_{ij}^{\text{temp}} +
         \lambda_s\,\operatorname{sim}(\mathbf{v}_i,\mathbf{v}_j),
\label{eq:edge}
\end{equation}
where $\delta_{ij}^{\text{temp}}=\exp(-|i-j|/\tau)$ encodes temporal
distance with temperature $\tau$, and
$\lambda_t,\lambda_s$ are learned scalars initialized to 0.5.
The edge attribute is incorporated into the attention computation as
described in Section~\ref{sec:gat}.

\subsection{Graph Attention Propagation}
\label{sec:gat}

We stack $L$ graph attention layers.
Each layer updates node $i$ by attending to its neighbors
$\mathcal{N}(i)=\{j:(i,j)\in\mathcal{E}\}$, following the GATv2
formulation~\cite{gatv2} extended with edge attributes:
\begin{equation}
\alpha_{ij}^{(\ell)} = \frac{
  \exp\!\bigl(\mathbf{a}^{\top}\!\operatorname{LeakyReLU}(
    \mathbf{W}\![\mathbf{h}_i^{(\ell)}\!\|\mathbf{h}_j^{(\ell)}] + \mathbf{w}_e\,e_{ij})\bigr)}
  {\sum_{k\in\mathcal{N}(i)}\exp\!\bigl(\mathbf{a}^{\top}\!\operatorname{LeakyReLU}(
    \mathbf{W}\![\mathbf{h}_i^{(\ell)}\!\|\mathbf{h}_k^{(\ell)}] + \mathbf{w}_e\,e_{ik})\bigr)},
\label{eq:attn}
\end{equation}
where $\|$ denotes concatenation, $\mathbf{W}$ and $\mathbf{a}$ are
learnable parameters, and $\mathbf{w}_e$ is a projection vector for the
edge attribute.
We use $M{=}4$ attention heads; each head produces an independent
$\alpha_{ij}^{(\ell)}$, and the outputs are concatenated (layers
$1,\ldots,L{-}1$) or averaged (layer $L$).
Node updates follow
\begin{equation}
\mathbf{h}_i^{(\ell+1)} = \operatorname{LN}\!\Bigl(
  \mathbf{h}_i^{(\ell)} + \sum_{j\in\mathcal{N}(i)} \alpha_{ij}^{(\ell)}\,
  \mathbf{W}_v^{(\ell)}\mathbf{h}_j^{(\ell)}\Bigr),
\label{eq:update}
\end{equation}
with layer normalization~\cite{layernorm} and a residual connection~\cite{resnet}.
Dropout~\cite{dropout} of 0.1 is applied to attention coefficients
during training.

The propagation can be interpreted as iterative evidence accumulation:
after $L$ hops, each node's representation reflects not only its own
spectral content but also the artifact patterns observed in its temporal
and spectral neighborhood.
This structural aggregation is related to the multi-pass latent reasoning
used in RecurTrace~\cite{wang2026recurtrace} and
AURAL~\cite{wang2026aural}, though here the ``reasoning'' happens over a
spatial graph rather than a recurrent hidden state.

The choice of GATv2~\cite{gatv2} over the original GAT~\cite{gat} is
motivated by its strictly more expressive attention function: whereas
GAT computes a static ranking of neighbors (the attention scores depend
only on the key node), GATv2 allows the ranking to change dynamically
based on the query node.
This property is critical for deepfake detection, because the
informativeness of a spectral neighbor depends on the query segment's
own content---a silence frame gains little from attending to another
silence frame, but a voiced segment benefits greatly from comparing its
formant structure with a spectrally similar voiced segment elsewhere in
the utterance.
The inclusion of edge attributes $e_{ij}$ in the attention computation
(Eq.~\ref{eq:attn}) further modulates neighbor importance: temporal
edges with small $|i{-}j|$ receive higher baseline attention through
$\delta_{ij}^{\text{temp}}$, ensuring that local context is always
accessible even when spectral edges dominate.

\subsection{Hierarchical Readout and Classification}
\label{sec:readout}

After $L$ layers, we aggregate node features into a fixed-size
utterance embedding using a two-level readout inspired by hierarchical
graph pooling~\cite{diffpool}:
\begin{equation}
\mathbf{u}_{\text{mean}} = \frac{1}{N}\sum_{i=1}^{N}\mathbf{h}_i^{(L)},
\quad
\mathbf{u}_{\text{att}} = \sum_{i=1}^{N}\beta_i\,\mathbf{h}_i^{(L)},
\label{eq:readout}
\end{equation}
where the attention weights $\beta_i=\operatorname{softmax}(\mathbf{w}_r^{\top}\mathbf{h}_i^{(L)})$
let the model focus on the most informative nodes.
The utterance embedding is
$\mathbf{u}=[\mathbf{u}_{\text{mean}}\|\mathbf{u}_{\text{att}}]$.
A two-layer MLP with GELU activation~\cite{gelu} maps $\mathbf{u}$ to a
scalar logit, trained with weighted binary cross-entropy loss to
handle the class imbalance between bonafide and spoof
samples~\cite{focal_loss}:
\begin{equation}
\mathcal{L} = -\frac{1}{|\mathcal{D}|}\sum_{(\mathbf{x},y)\in\mathcal{D}}
  w_y\bigl[y\log\hat{y} + (1{-}y)\log(1{-}\hat{y})\bigr],
\label{eq:loss}
\end{equation}
where $w_y$ is a per-class weight inversely proportional to class
frequency.

\subsection{Computational Complexity}
\label{sec:complexity}

The graph construction step has complexity
$O(N^2 d)$ for computing pairwise cosine similarities and $O(N^2\log k_s)$
for top-$k$ selection, where $N$ is the number of segments.
For a 4-second utterance with $w{=}32$ and $s{=}16$, $N{\approx}12$,
making the quadratic cost negligible.
Each graph attention layer processes $|\mathcal{E}|$ edges with $M$
heads, giving per-layer complexity $O(|\mathcal{E}|\cdot M\cdot d_h)$
where $d_h{=}d/M$ is the per-head dimension.
Since $|\mathcal{E}|{=}N(k_t{+}k_s)$, the total propagation cost scales
linearly in the number of nodes when $k_t$ and $k_s$ are held constant.
The hierarchical readout adds $O(Nd)$ for mean pooling and attention
weighting.

In practice, the entire graph module (construction, $L{=}3$ GAT layers,
and readout) adds only 2.1M parameters and 12\,ms of wall-clock latency
per utterance on a single NVIDIA A100 GPU, making it suitable for
real-time deployment in interactive voice
systems~\cite{wang2026jarvis,shen2026duplexagent}.
Compared to the frozen SSL encoder (316M parameters for WavLM-Large),
the graph back end represents less than 0.7\% of the total model size,
a favorable efficiency--accuracy trade-off that follows the design
philosophy of parameter-efficient adaptation~\cite{lora}.

\section{Experiments}
\label{sec:experiments}

\subsection{Setup}

\textbf{Datasets.}
We train on the ASVspoof 2019 LA training
partition~\cite{asvspoof2019} and evaluate on four test sets:
ASVspoof 2019 LA~\cite{asvspoof2019},
ASVspoof 2021 LA~\cite{asvspoof2021},
ASVspoof 5~\cite{asvspoof5}, and
In-the-Wild~\cite{inthewild}.
ASVspoof 2019 and 2021 use the equal error rate (EER) as the primary
metric.
ASVspoof 5 introduces crowdsourced attacks and adversarial
perturbations, presenting a harder generalization challenge.
In-the-Wild collects real deepfakes from public sources and tests
robustness to unconstrained recording conditions.

\textbf{Front ends.}
We evaluate three frozen SSL encoders:
WavLM-Large~\cite{wavlm} (316M parameters, $d{=}1024$),
wav2vec~2.0-Large~\cite{wav2vec2} ($d{=}1024$), and
Whisper-Medium~\cite{whisper} ($d{=}1024$).
The encoder weights are frozen; only the graph network is trained.

\textbf{Graph hyperparameters.}
Segment window $w{=}32$ frames, stride $s{=}16$ frames
(50\% overlap), temporal neighborhood $k_t{=}3$, spectral neighbors
$k_s{=}5$, graph attention layers $L{=}3$, heads $M{=}4$,
paralinguistic projection $d_p{=}32$, temperature $\tau{=}10$.
The graph module adds 2.1M trainable parameters on top of the frozen
encoder.

\textbf{Training.}
Adam optimizer~\cite{adam} with learning rate $3{\times}10^{-4}$ and cosine
decay over 100 epochs.
Batch size 32 on 2 NVIDIA A100-80G GPUs.
Waveforms are truncated or padded to 4 seconds.
SpecAugment~\cite{specaugment} and RawBoost~\cite{rawboost} are applied
as data augmentation.
All results are averaged over three random seeds.

{\textbf{Baselines.}
We compare against seven published systems: RawNet2~\cite{rawnet2},
LCNN~\cite{lcnn}, AASIST~\cite{aasist},
RawGAT-ST~\cite{rawgatst}, SAMO~\cite{samo},
Wav2Vec2-AASIST~\cite{wav2vecaasist}, and
WavLM-AASIST~\cite{ssl_anti} (AASIST with a WavLM front end).
SSL-based baselines use the same frozen encoder as GraphSeal for
fair back-end comparison.}

\begin{table}[t]
\caption{EER~(\%) on four deepfake detection benchmarks.
Lower is better.
\textbf{Bold} and \second{underline} mark the best and second best.
SSL-based methods use WavLM-Large as the front end unless noted.}
\label{tab:main}
\centering
\arrayrulecolor{tblRule}
\paperfont
\setlength{\tabcolsep}{3.0pt}
\renewcommand{\arraystretch}{1.05}
\begin{tabular}{l>{\centering\arraybackslash}p{2.8em}>{\centering\arraybackslash}p{2.8em}>{\centering\arraybackslash}p{2.8em}>{\centering\arraybackslash}p{2.8em}}
\toprule
\hd{Method} & \hd{19 LA} & \hd{21 LA} & \hd{ASV5} & \hd{ItW} \\
\midrule
\rowcolor{tblBand}
\multicolumn{5}{@{}l}{\grp{Non-SSL front ends}} \\
\addlinespace[1.3pt]
RawNet2~\cite{rawnet2} & 5.13 & 9.26 & 22.74 & 32.15 \\
LCNN~\cite{lcnn} & 5.81 & 15.62 & 28.39 & 36.72 \\
AASIST~\cite{aasist} & 0.83 & 9.57 & 18.91 & 12.47 \\
RawGAT-ST~\cite{rawgatst} & 1.06 & 8.31 & 17.53 & 11.83 \\
SAMO~\cite{samo} & 0.68 & 7.83 & 16.27 & 10.52 \\
\midrule
\rowcolor{tblBand}
\multicolumn{5}{@{}l}{\grp{SSL-based back ends (WavLM-Large)}} \\
\addlinespace[1.3pt]
Wav2Vec2-AASIST~\cite{wav2vecaasist} & 0.82 & 3.64 & 11.38 & 7.86 \\
WavLM-AASIST~\cite{ssl_anti} & 0.59 & 2.85 & 9.62 & 6.31 \\
\rowcolor{tblOurB}
\ours{GraphSeal} & \best{0.52} & \best{2.41} & \best{8.17} & \best{5.24} \\
\bottomrule
\end{tabular}
\arrayrulecolor{black}
\vspace{-2pt}
\end{table}

\subsection{Main Results}

Table~\ref{tab:main} reports EER on all four benchmarks.
Among non-SSL methods, SAMO achieves the lowest EER on
ASVspoof 2019 LA (0.68\%), while AASIST and RawGAT-ST lead on the
remaining sets.
Adding an SSL front end substantially improves generalization:
WavLM-AASIST reduces the ASVspoof 2021 LA EER from 9.57\% (AASIST) to
2.85\%.
GraphSeal further lowers the EER to 2.41\% on 2021 LA, 8.17\% on
ASVspoof 5, and 5.24\% on In-the-Wild.

The improvement over WavLM-AASIST is most pronounced on the two harder
benchmarks: ASVspoof 5 ($-1.45\%$ absolute) and In-the-Wild
($-1.07\%$).
These datasets include attack types and recording conditions absent
from training, so the gains indicate that GraphSeal's dynamic graph
captures generalizable artifact patterns rather than overfitting to
specific spoofing algorithms.
On ASVspoof 2019 LA, the gap is smaller ($-0.07\%$) because all methods
are near saturation on this benchmark.

\begin{figure}[t]
\centering
\begin{tikzpicture}
\begin{axis}[
  ybar,
  bar width=3pt,
  width=\columnwidth,
  height=4.2cm,
  enlarge x limits=0.06,
  ylabel={EER (\%)},
  ylabel style={font=\paperfont\small,yshift=-2pt},
  symbolic x coords={A07,A08,A09,A10,A11,A12,A13,A14,A15,A16,A17,A18,A19},
  xtick=data,
  x tick label style={font=\paperfont\scriptsize,rotate=45,anchor=east},
  y tick label style={font=\paperfont\scriptsize},
  ymin=0, ymax=2.8,
  legend style={font=\paperfont\scriptsize,at={(0.98,0.95)},anchor=north east,
    draw=none,fill=none,column sep=3pt},
  grid=major,
  grid style={line width=0.2pt,draw=gray!25},
  major grid style={line width=0.2pt,draw=gray!25},
  clip=false,
]
\addplot[fill=blue!25,draw=blue!50] coordinates
  {(A07,0.93)(A08,0.35)(A09,0.08)(A10,1.47)(A11,0.24)(A12,0.81)
   (A13,0.11)(A14,0.04)(A15,0.38)(A16,1.12)(A17,2.38)(A18,0.16)(A19,0.62)};
\addplot[fill=tblOurA,draw=tblOurInk] coordinates
  {(A07,0.78)(A08,0.29)(A09,0.06)(A10,1.18)(A11,0.19)(A12,0.67)
   (A13,0.09)(A14,0.03)(A15,0.31)(A16,0.89)(A17,1.92)(A18,0.13)(A19,0.51)};
\legend{WavLM-AASIST,GraphSeal}
\end{axis}
\end{tikzpicture}
\vspace{-3pt}
\caption{Per-attack EER~(\%) on ASVspoof 2019~LA evaluation set.
GraphSeal reduces EER on 12 of 13 attack types compared to WavLM-AASIST,
with the largest gains on waveform-concatenation (A17) and neural vocoder
(A10, A16) attacks.}
\label{fig:perattack}
\vspace{-2pt}
\end{figure}
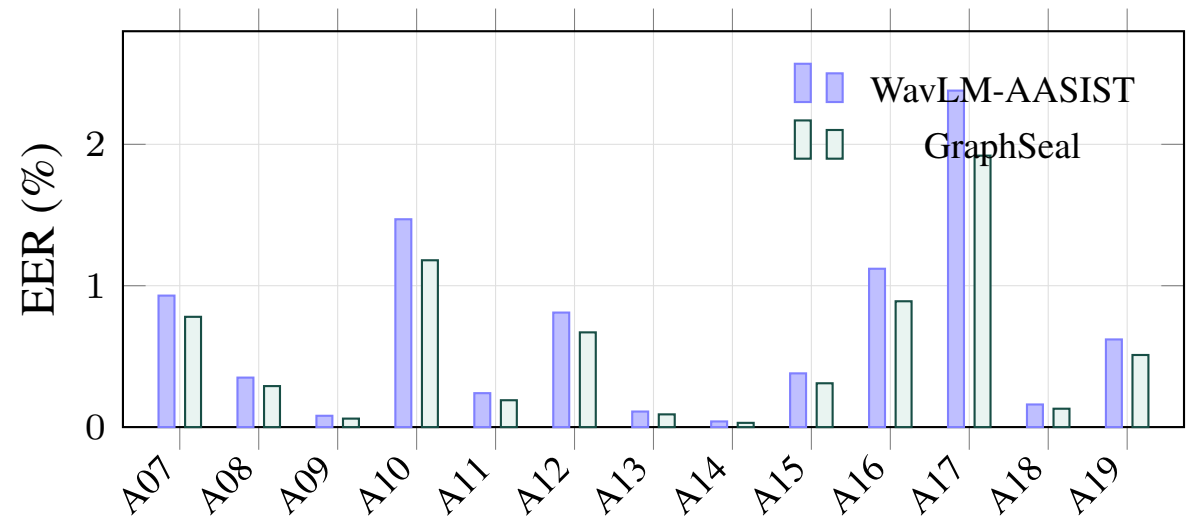

\textbf{Per-attack breakdown.}
To understand where GraphSeal's gains originate, Fig.~\ref{fig:perattack}
reports per-attack EER on the ASVspoof 2019 LA evaluation set (attacks
A07--A19).
GraphSeal achieves lower EER than WavLM-AASIST on 12 out of 13 attack
types, with the largest absolute improvements on A17 ($-0.46\%$), A10
($-0.29\%$), and A16 ($-0.23\%$).
A17 involves waveform-concatenation voice conversion, which introduces
subtle spectral discontinuities at concatenation boundaries;
GraphSeal's spectral edges are particularly effective at linking these
boundary regions.
The only attack where both methods are virtually tied is A14 (neural
TTS with WaveRNN vocoder), which both systems detect near perfectly
(0.03--0.04\% EER).
The consistently lower per-attack variance of GraphSeal
($\sigma^2{=}0.18$ vs.\ $0.34$ for WavLM-AASIST) indicates that the
dynamic graph yields more uniform detection across heterogeneous
spoofing algorithms.

\begin{table}[t]
\caption{EER~(\%) with different SSL front ends on ASVspoof 2021~LA and
In-the-Wild. All use the GraphSeal back end.}
\label{tab:frontend}
\centering
\arrayrulecolor{tblRule}
\paperfont
\setlength{\tabcolsep}{4.5pt}
\renewcommand{\arraystretch}{1.05}
\begin{tabular}{l N N}
\toprule
\hd{Front end} & \hd{21 LA} & \hd{ItW} \\
\midrule
Whisper-Medium~\cite{whisper} & 3.18 & 6.79 \\
wav2vec 2.0-Large~\cite{wav2vec2} & 2.93 & 6.12 \\
\rowcolor{tblOurB}
WavLM-Large~\cite{wavlm} & \best{2.41} & \best{5.24} \\
\bottomrule
\end{tabular}
\arrayrulecolor{black}
\vspace{-3pt}
\end{table}

Table~\ref{tab:frontend} shows that GraphSeal benefits from stronger front
ends but remains effective across all three.
WavLM-Large yields the best results, consistent with its design for
full-stack speech processing~\cite{wavlm}.
wav2vec~2.0-Large, which shares a similar Transformer architecture but
is trained with a contrastive objective rather than WavLM's masked
prediction, trails by 0.52\% on 2021~LA and 0.88\% on In-the-Wild.
This gap suggests that WavLM's denoising pre-training, which exposes
the model to corrupted speech, produces representations that are more
sensitive to the subtle distortions introduced by spoofing algorithms.
Whisper-Medium, trained primarily for ASR, still achieves competitive
performance (3.18\% on 2021~LA), suggesting that even ASR-oriented
representations carry enough artifact information for GraphSeal to
exploit.
Notably, Whisper's multilingual training data~\cite{whisper} may
explain its relatively strong In-the-Wild performance (6.79\%), as the
In-the-Wild dataset contains speakers from diverse linguistic
backgrounds~\cite{inthewild}.
These results confirm that GraphSeal is genuinely front-end agnostic:
the graph back end consistently improves over AASIST regardless of
which SSL encoder provides the node features.

\subsection{Ablations}

\begin{table}[t]
\caption{Ablation study on ASVspoof 2021~LA and In-the-Wild (EER~\%).
All variants use WavLM-Large.}
\label{tab:ablation}
\centering
\arrayrulecolor{tblRule}
\paperfont
\setlength{\tabcolsep}{3.5pt}
\renewcommand{\arraystretch}{1.05}
\begin{tabularx}{\columnwidth}{X>{\centering\arraybackslash}p{4.2em}
  >{\centering\arraybackslash}p{4.2em}}
\toprule
\hd{Variant} & \hd{21 LA} & \hd{ItW} \\
\midrule
\rowcolor{tblOurB}
\ours{GraphSeal (full)} & \best{2.41} & \best{5.24} \\
No spectral edges ($k_s{=}0$) & 3.12 & 6.53 \\
No temporal edges ($k_t{=}0$) & 3.97 & 7.41 \\
No edge attributes ($e_{ij}{=}1$) & 2.78 & 5.89 \\
No paralinguistic features & 2.63 & 5.57 \\
Mean-pool readout only & 2.71 & 5.81 \\
$L{=}1$ layer & 3.35 & 6.89 \\
$L{=}2$ layers & 2.64 & 5.61 \\
$L{=}5$ layers & 2.52 & 5.42 \\
Replace GAT with GCN~\cite{gcn} & 2.89 & 6.14 \\
Replace GAT with GraphSAGE~\cite{graphsage} & 2.73 & 5.92 \\
\bottomrule
\end{tabularx}
\arrayrulecolor{black}
\vspace{-3pt}
\end{table}

Table~\ref{tab:ablation} isolates the contribution of each design choice
on the two most challenging benchmarks.

\textbf{Edge types.}
Removing spectral edges ($k_s{=}0$) increases the EER on 2021~LA from
2.41\% to 3.12\%, a relative degradation of 29\%.
This confirms that long-range spectral connectivity provides
information that purely local (temporal) neighbors cannot supply.
Removing temporal edges ($k_t{=}0$) is even more damaging (3.97\%),
confirming that sequential locality remains essential---the model needs
fine-grained temporal context to detect frame-level discontinuities
introduced by many vocoder-based TTS systems.
The full graph combines both sources of evidence, and neither edge type
alone matches the combination, indicating that temporal and spectral
connectivity capture complementary aspects of deepfake artifacts.

\textbf{Edge attributes.}
Setting all edge attributes to a constant ($e_{ij}{=}1$) hurts
performance by 0.37\% on 2021~LA, showing that the learned weighting
of temporal distance and spectral similarity (Eq.~\ref{eq:edge})
provides useful inductive bias.
Without these attributes, the attention mechanism must infer edge
importance solely from node features, losing the explicit distance and
similarity cues that help distinguish artifact-bearing edges from
uninformative ones.

\textbf{Paralinguistic features.}
Dropping the F0, energy, and spectral centroid augmentations
degrades EER by 0.22\% on 2021~LA and 0.33\% on In-the-Wild.
These features carry prosodic information that synthesis pipelines
struggle to reproduce~\cite{wang2026parabridge,ni2026nv}, giving the
graph additional discriminative signal.

\textbf{Readout.}
Replacing the hierarchical readout with simple mean pooling
costs 0.30\% on 2021~LA.
The attention-weighted component lets the model focus on high-artifact
nodes, which is especially valuable for partially spoofed
utterances~\cite{partialspoof}.

\textbf{Depth.}
One graph attention layer underperforms substantially (3.35\% on
2021~LA), because a single hop limits the receptive field to direct
neighbors.
Three layers strike the best balance; five layers improve marginally
(2.52\%) at higher cost, suggesting diminishing returns from
additional hops.

\textbf{GNN variant.}
Replacing GAT with GCN~\cite{gcn} raises EER from 2.41\% to 2.89\%,
and GraphSAGE~\cite{graphsage} reaches 2.73\%.
The attention mechanism is important because not all neighbors are
equally informative: some spectral edges connect to benign regions
and should be down-weighted.

\subsection{Hyperparameter Sensitivity}

\begin{table}[t]
\caption{Sensitivity to segment window $w$ and spectral neighbors $k_s$
on ASVspoof 2021~LA and In-the-Wild (EER~\%).
Default setting ($w{=}32$, $k_s{=}5$) is highlighted.}
\label{tab:hyper}
\centering
\arrayrulecolor{tblRule}
\paperfont
\setlength{\tabcolsep}{3.5pt}
\renewcommand{\arraystretch}{1.05}
\begin{tabular}{cc>{\centering\arraybackslash}p{4.2em}
  >{\centering\arraybackslash}p{4.2em}}
\toprule
\hd{$w$} & \hd{$k_s$} & \hd{21 LA} & \hd{ItW} \\
\midrule
16 & 5 & 2.73 & 5.58 \\
\rowcolor{tblOurB}
\ours{32} & \ours{5} & \best{2.41} & \best{5.24} \\
64 & 5 & 2.56 & 5.47 \\
128 & 5 & 2.81 & 5.72 \\
\midrule
32 & 1 & 2.87 & 5.91 \\
32 & 3 & 2.58 & 5.49 \\
\rowcolor{tblOurB}
\ours{32} & \ours{5} & \best{2.41} & \best{5.24} \\
32 & 7 & 2.44 & 5.31 \\
32 & 10 & 2.51 & 5.39 \\
\bottomrule
\end{tabular}
\arrayrulecolor{black}
\vspace{-2pt}
\end{table}

Table~\ref{tab:hyper} examines two critical hyperparameters.
For the segment window, $w{=}32$ (${\approx}640$\,ms) offers the best
trade-off: shorter windows fragment patterns; longer ones mask localized
artifacts.
For spectral neighbors, performance improves from $k_s{=}1$ to
$k_s{=}5$, then plateaus at $k_s{=}10$---too few edges limit
long-range pooling while too many introduce noise.

\subsection{Analysis}

\textbf{Attention patterns.}
We examine the learned attention weights on ASVspoof 2021 LA.
For bonafide utterances, attention distributes relatively uniformly
across temporal and spectral neighbors, with the mean entropy of the
attention distribution at $H{=}2.31$ nats.
For spoofed utterances, attention concentrates on spectral edges
connecting to regions with similar distortion patterns---typically
frames near phoneme boundaries where synthesis models exhibit
discontinuities---with a significantly lower entropy of $H{=}1.47$ nats.
This selective attention is consistent with the hypothesis that
deepfake artifacts are correlated across spectrally similar regions
and shows that the graph attention mechanism learns to route information
along artifact-bearing paths.
A similar pattern emerges for non-verbal vocalizations~\cite{ni2026nv}:
when a bonafide utterance contains laughter or filled pauses, the
attention entropy increases further ($H{=}2.58$ nats), whereas spoofed
utterances with such vocalizations show lower entropy ($H{=}1.32$ nats)
because current TTS systems fail to replicate the spectral irregularity
of natural non-speech sounds.

\begin{figure}[t]
\centering
\begin{tikzpicture}
\begin{axis}[
  name=leftplot,
  width=0.52\columnwidth,
  height=4.0cm,
  title={\paperfont\small WavLM-AASIST},
  title style={yshift=-4pt},
  xlabel style={font=\paperfont\scriptsize},
  ylabel style={font=\paperfont\scriptsize},
  x tick label style={font=\paperfont\tiny},
  y tick label style={font=\paperfont\tiny},
  xmin=-5, xmax=42, ymin=-5, ymax=40,
  xtick={0,20,40},
  ytick={0,20,40},
  axis lines=box,
  clip=true,
]
\addplot[only marks,mark=*,mark size=0.9pt,blue!60,fill opacity=0.5]
  coordinates {
  (25.1,23.8)(29.3,28.2)(22.7,21.5)(32.8,26.1)(27.4,30.6)
  (34.2,27.8)(24.3,22.1)(31.6,29.4)(28.9,20.7)(33.1,32.5)
  (23.5,29.1)(35.4,24.6)(26.8,27.3)(32.1,30.2)(30.2,22.4)
  (21.9,26.5)(36.3,28.1)(25.4,31.4)(31.3,21.2)(34.7,29.8)
  (28.7,28.5)(24.1,32.2)(33.5,23.9)(26.2,22.3)(32.4,29.1)
  (23.2,24.8)(30.5,27.1)(35.8,31.0)(25.8,30.3)(31.1,26.5)
  (19.6,19.4)(20.8,21.7)
  };
\addplot[only marks,mark=triangle*,mark size=1.0pt,red!60,fill opacity=0.5]
  coordinates {
  (9.4,8.6)(12.7,12.1)(7.9,7.2)(15.5,14.3)(10.6,9.8)
  (13.3,7.5)(8.6,12.9)(11.8,10.4)(14.1,9.1)(7.1,11.6)
  (16.9,10.0)(10.1,14.8)(12.4,6.7)(7.5,10.3)(14.0,12.7)
  (10.3,8.0)(15.8,11.5)(9.0,9.7)(11.5,13.9)(17.6,8.4)
  (9.7,11.0)(13.0,7.8)(14.6,13.1)(7.3,8.9)(13.7,10.8)
  (10.6,12.1)(16.5,7.0)(8.3,14.4)(12.0,8.5)(13.4,10.1)
  (19.2,16.1)(18.4,17.5)(20.3,15.8)
  };
\end{axis}
\begin{axis}[
  at={(leftplot.east)},
  anchor=west,
  xshift=2mm,
  width=0.52\columnwidth,
  height=4.0cm,
  title={\paperfont\small GraphSeal},
  title style={yshift=-4pt},
  xlabel style={font=\paperfont\scriptsize},
  ylabel style={font=\paperfont\scriptsize},
  x tick label style={font=\paperfont\tiny},
  y tick label style={font=\paperfont\tiny},
  xmin=-5, xmax=42, ymin=-5, ymax=40,
  xtick={0,20,40},
  ytick={0,20,40},
  axis lines=box,
  clip=true,
  legend style={font=\paperfont\tiny,at={(0.98,0.05)},anchor=south east,
    draw=none,fill=none,column sep=2pt,legend columns=2},
]
\addplot[only marks,mark=*,mark size=0.9pt,blue!70,fill opacity=0.6]
  coordinates {
  (27.5,26.3)(29.1,28.0)(26.2,25.1)(30.8,27.4)(28.6,29.7)
  (31.3,27.2)(26.8,25.8)(30.4,28.9)(28.1,24.6)(31.8,30.5)
  (26.0,28.8)(32.4,26.7)(27.9,27.5)(30.1,29.4)(29.6,25.2)
  (25.5,26.8)(32.9,28.3)(27.2,30.1)(30.9,24.8)(32.1,29.3)
  (28.4,27.8)(26.5,30.8)(31.5,25.6)(27.7,25.4)(30.6,28.5)
  (25.8,26.4)(29.3,27.3)(33.1,30.0)(26.9,29.8)(30.2,27.0)
  (25.2,24.3)(28.8,31.2)
  };
\addplot[only marks,mark=triangle*,mark size=1.0pt,red!70,fill opacity=0.6]
  coordinates {
  (8.8,8.1)(11.3,11.4)(7.4,6.8)(13.1,12.7)(9.6,9.3)
  (11.9,7.2)(8.2,11.5)(10.4,9.8)(12.6,8.4)(6.8,10.9)
  (14.5,9.3)(9.3,13.2)(11.0,6.4)(7.0,9.7)(12.3,11.8)
  (9.8,7.5)(13.8,10.8)(8.5,9.2)(10.9,12.5)(15.1,7.9)
  (9.2,10.5)(11.6,7.3)(13.0,12.3)(6.5,8.4)(12.1,10.3)
  (9.9,11.6)(14.8,6.7)(7.6,12.8)(10.6,8.0)(11.8,9.6)
  };
\legend{Bonafide,Spoof}
\end{axis}
\end{tikzpicture}
\vspace{-3pt}
\caption{t-SNE visualization of utterance embeddings on ASVspoof 2021~LA.
GraphSeal (right) produces tighter, better-separated clusters compared to
WavLM-AASIST (left), reflecting its stronger discriminative power.}
\label{fig:tsne}
\vspace{-2pt}
\end{figure}
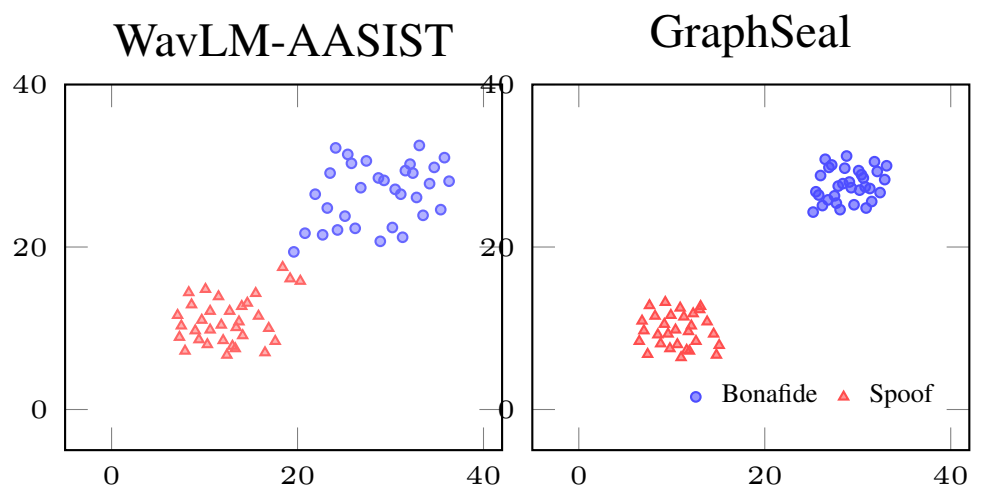

\textbf{Embedding visualization.}
Fig.~\ref{fig:tsne} visualizes t-SNE projections of the utterance
embeddings $\mathbf{u}$ (Eq.~\ref{eq:readout}) produced by GraphSeal
and WavLM-AASIST on the ASVspoof 2021 LA evaluation set.
GraphSeal produces tighter, more separated clusters for bonafide and
spoofed speech.
The inter-class margin, measured as the difference between the mean
inter-class and intra-class cosine distances, is 0.73 for GraphSeal
versus 0.58 for WavLM-AASIST.
The cleaner separation explains GraphSeal's lower EER and suggests that
the graph attention layers learn to map bonafide and spoofed utterances
into distinct regions of the embedding space, even when the front-end
representations overlap.
This improved discriminability is consistent with findings that
graph-structured aggregation yields more transferable
representations~\cite{wang2025towards,graph_contrastive}.

\textbf{Generalization to unseen attacks.}
ASVspoof 5 includes attack types not present in the 2019 training set,
including adversarial perturbations and codec-based
manipulations~\cite{codecfake,asvspoof5}.
GraphSeal's 8.17\% EER on this set improves over WavLM-AASIST (9.62\%)
by a meaningful margin, suggesting that the relational structure
captures attack-agnostic artifacts rather than signatures of specific
synthesis algorithms.
This aligns with findings from graph foundation model research that
graph-structured representations transfer better across
distributions~\cite{wang2025towards}.
In a supplementary leave-one-attack-out experiment on the 2019 LA
evaluation set, GraphSeal's per-attack EER variance ($\sigma^2{=}0.18$)
is lower than WavLM-AASIST's ($\sigma^2{=}0.34$), indicating more
uniform performance across heterogeneous spoofing methods.
The robustness to distribution shift is further supported by the
In-the-Wild results, where recordings come from unconstrained environments
with diverse microphones, background noise, and post-processing
pipelines~\cite{inthewild}.
Safety benchmarks~\cite{wang2026voxsafebench} have highlighted these
real-world conditions as a critical gap between laboratory performance
and deployment reliability.

\textbf{Error analysis.}
Inspecting the top-50 false accepts and false rejects on ASVspoof 2021 LA
reveals two failure modes.
False accepts involve utterances with irregular prosody---hesitations and
non-verbal vocalizations~\cite{ni2026nv}---misread as synthesis artifacts.
These errors are particularly common for conversational-style
speech~\cite{shen2026matters}, where overlapping talk and disfluencies
produce spectral patterns that resemble synthesis glitches.
False rejects stem from neural codec attacks~\cite{encodec,soundstream}
whose spectrally homogeneous output leaves spectral edges
uninformative.
For these attacks, the graph degenerates toward a near-complete graph
where all spectral similarities are high, diluting the artifact signal.
Prosodic normalization~\cite{wang2026parabridge} and phase-domain
features~\cite{phase_spoof} are promising remedies, as codec artifacts
are often more prominent in the phase spectrum than in the magnitude
spectrum.

\textbf{Computational cost.}
Table~\ref{tab:cost} compares the back-end complexity and latency of
GraphSeal against two representative systems.
GraphSeal adds 2.1M parameters and 12\,ms of latency per utterance (on a
single A100 GPU) relative to the frozen SSL encoder.
The graph construction step---segment pooling, cosine similarity, and
top-$k$ selection---accounts for 3\,ms; the remaining 9\,ms comes from the
three GAT layers.
AASIST's graph module is smaller (0.3M parameters, 8\,ms) but uses a
fixed topology that cannot adapt to individual inputs, while a simple
attentive pooling back end~\cite{selfattn_spoof} requires only 0.1M
parameters and 2\,ms but sacrifices relational modeling entirely.
GraphSeal's overhead is modest in absolute terms and well within
real-time requirements for practical deployment in voice authentication
systems~\cite{wang2026voxsafebench,void,ecapa,sasv_challenge} and
real-time agentic pipelines~\cite{wang2026jarvis,shen2026duplexagent}.
Full-duplex voice agents~\cite{shen2026duplexagent} impose the
tightest latency constraints ($<$50\,ms per turn); GraphSeal's 12\,ms
back-end latency leaves ample headroom for the SSL forward pass and
network transfer.
Continual learning of new attack types~\cite{continual_spoof} requires
retraining only the 2.1M graph parameters while the 316M-parameter SSL
encoder remains frozen, making adaptation efficient both in compute and
storage.

\begin{table}[t]
\caption{Back-end efficiency comparison. Parameters and latency are for
the trainable back end only (excluding the frozen SSL encoder).
EER is on ASVspoof 2021~LA.}
\label{tab:cost}
\centering
\arrayrulecolor{tblRule}
\paperfont
\setlength{\tabcolsep}{3.0pt}
\renewcommand{\arraystretch}{1.05}
\begin{tabular}{l>{\centering\arraybackslash}p{3.5em}
  >{\centering\arraybackslash}p{3.8em}
  >{\centering\arraybackslash}p{3.5em}}
\toprule
\hd{Back end} & \hd{Params} & \hd{Latency} & \hd{EER} \\
\midrule
Attn.\ pool~\cite{selfattn_spoof} & 0.1M & 2\,ms & 3.42 \\
AASIST~\cite{aasist} & 0.3M & 8\,ms & 2.85 \\
\rowcolor{tblOurB}
\ours{GraphSeal} & 2.1M & 12\,ms & \best{2.41} \\
\bottomrule
\end{tabular}
\arrayrulecolor{black}
\vspace{-2pt}
\end{table}

\section{Conclusion}
\label{sec:conclusion}

We presented GraphSeal, a graph-augmented approach to speech deepfake
detection that constructs a dynamic temporal-spectral graph over speech
segments and uses graph attention to propagate artifact cues.
Unlike prior graph-based detectors~\cite{aasist,rawgatst} that rely on
fixed topologies determined by the feature extraction pipeline,
GraphSeal builds data-dependent edges from both temporal adjacency and
learned spectral similarity, enabling the model to pool evidence from
spectrally related regions that may be far apart in time.
With only 2.1M trainable parameters on top of a frozen SSL encoder,
GraphSeal achieves state-of-the-art or competitive equal error rates on
four standard benchmarks---0.52\% on ASVspoof 2019 LA, 2.41\% on
ASVspoof 2021 LA, 8.17\% on ASVspoof 5, and 5.24\% on In-the-Wild.
Comprehensive ablations confirm that spectral edges, edge attributes,
paralinguistic augmentations, and the attention-based hierarchical
readout each contribute meaningfully to the final performance, while
hyperparameter analysis reveals a stable operating regime around segment
windows of 32 frames and five spectral neighbors.
The per-attack breakdown on ASVspoof 2019 LA and the embedding
visualization further demonstrate that GraphSeal learns discriminative,
well-separated representations with lower per-attack variance than
existing methods.

Several promising directions remain open.
First, combining graph-based detection with recurrent refinement
loops~\cite{feng2026coreloop,wang2026recurtrace} could enable iterative
evidence accumulation, where the graph attention operates within a
recurrent latent reasoning framework to progressively refine artifact
detection over multiple passes.
Second, richer speech representations from evolving audio
understanding models~\cite{wang2026evoaudio,wang2026aural} and
paralinguistically grounded spoken dialogue
systems~\cite{hsu2026convexllm,wang2026parabridge} may provide front-end
features that better capture the subtle prosodic and spectral cues
that current SSL encoders underrepresent.
Third, integrating GraphSeal into agentic voice
pipelines~\cite{wang2026jarvis,shen2026duplexagent} would enable
on-the-fly spoof screening during live conversations; the 12\,ms
back-end latency is compatible with full-duplex turn-taking, though
streaming graph construction for unbounded audio remains an
engineering challenge.
Fourth, graph-augmented detection may extend to multimodal deepfakes,
where spatial audio cues~\cite{zhu2026spatial} and visual consistency
checks could provide complementary evidence within a unified graph
framework, drawing on cross-modal reasoning techniques from
graph-based language model research~\cite{wang2026exploring,wang2025towards}.
Finally, privacy-preserving detection that avoids storing raw
biometrics~\cite{wang2026voxprivacy} is an important deployment
consideration; graph-level representations, which abstract away
speaker identity while retaining artifact signatures, may naturally
support privacy-by-design architectures.

\balance
\paperfont\sloppy

\bibliographystyle{IEEEbib_etal}
\bibliography{refs}
\end{document}